\documentclass{amsart}

\newcommand{\rmi}{\textrm i}

\newtheorem{theorem}{Theorem}[section]
\newtheorem{definition}{Definition}[section]
\newtheorem{proposition}{Proposition}[section]
\newtheorem{remark}{Remark}[section]

\makeatletter
\CheckCommand*{\@cite}[2]{%
  {%
    \@citestyle[\citeform{#1}\if@tempswa, #2\fi]%
  }%
}
\renewcommand*{\@cite}[2]{%
  {%
    \@citestyle\citeform{#1}\if@tempswa, #2\fi
  }%
}
\makeatother

\author{Vedad Pasic}
\address{
Department of Mathematics \newline \indent
Faculty of Science and Mathematics \newline \indent
University of Tuzla  \newline \indent
Bosnia and Herzegovina}
\email{vedad.pasic@untz.ba}
\urladdr{http://pmf.untz.ba/vedad/}
\author{Elvis Barakovic}
\address{
Department of Mathematics \newline \indent
Faculty of Science and Mathematics \newline \indent
University of Tuzla  \newline \indent
Bosnia and Herzegovina}
\email{elvis.barakovic@untz.ba}
\urladdr{http://pmf.untz.ba/staff/elvis.barakovic/}

\title{Axial torsion waves in metric-affine gravity}


\begin{document}
\begin{abstract}
We construct new explicit vacuum solutions of quadratic metric-affine gravity. The approach of metric-affine gravity in using an independent affine connection produces a theory with 10+64 unknowns, which implies admitting torsion and possible nonmetricity. Our spacetimes are generalisations of classical pp-waves, four-dimensional Lorentzian spacetimes which admit a nonvanishing parallel spinor field. We generalize this definition to metric compatible spacetimes with pp-metric and purely axial torsion. It has been suggested that one can interpret that the axial component of torsion as the Hodge dual of the electromagnetic vector potential. We compare these solutions with our previous results and other solutions of classical models describing the interaction of gravitational and neutrino fields.
\end{abstract}
\maketitle

\section{Introduction}
Spacetime is considered to be a connected real 4-manifold $M$ equipped with a Lorentzian metric $g$
and an affine connection $\Gamma$. Our unknowns are the 10 independent
components of the metric tensor $g_{\mu\nu}$ and the 64 connection coefficients ${\Gamma^\lambda}_{\mu\nu}$.
This alternative theory of gravity is called \emph{metric-affine gravity}.
In \emph{quadratic} metric-affine gravity (QMAG), we define our action as
\begin{equation} \label{action}
S:=\int q(R),
\end{equation}
where $q$ is an $O(1, 3)$-invariant quadratic form on curvature $R$. Independently varying the action (\ref{action}) with respect to the metric $g$ and the connection $\Gamma$ produces the system of Euler-Lagrange equations which we will write symbolically as
\begin{eqnarray}
\label{eulerlagrangemetric}
\partial S/\partial g&=&0\\
\label{eulerlagrangeconnection}
\partial S/\partial \Gamma&=&0.
\end{eqnarray}
Our objective is the study of the system  \eqref{eulerlagrangemetric}, \eqref{eulerlagrangeconnection} and our motivation comes from Yang-Mills theory. The Yang-Mills action for the affine connection is a special case of (\ref{action})  where
\begin{equation}\label{YMcase}
q(R)=R^\kappa{}_{\lambda\mu\nu} R^\lambda{}_{\kappa}{}^{\mu\nu}.
\end{equation}
The origins of this theory lie in the works of \' Elie Cartan, Arthur Eddington, Albert Einstein, Erwin Schr\" odinger, Tullio Levi-Civita and Hermann Weyl.
The motivation for choosing a model of gravity which is purely quadratic
in curvature comes from Weyl, see end of Ref.~\cite{weyl1919eine}, and is explained in detail in our previous work.\textsuperscript{\cite{vassiliev2005quadratic,pasic2014pp}}
The results in this paper rely on the work of C.N. Yang and E.W. Mielke, who respectively showed that Einstein spaces satisfy equations (\ref{eulerlagrangeconnection}) and (\ref{eulerlagrangemetric}) in the Yang-Mills case (\ref{YMcase}).\textsuperscript{\cite{yang1974integral,mielke1981pseudoparticle}}
There are many works devoted to the study of the system (\ref{eulerlagrangemetric}), (\ref{eulerlagrangeconnection}) in the Yang-Mills case (\ref{YMcase}) and one can get an idea of the historical development of this theory of gravity from the references stated in Ref.~\cite{pasic2014pp}. Detailed descriptions of the irreducible pieces of curvature and quadratic forms on curvature can be found in Refs.~\cite{vassiliev2005quadratic,vassiliev2002pseudoinstantons} and our previous work in this theory can be found in Refs.~\cite{pasic2014pp,pasic2010new,pasic2014new,pasic2005pp}, where more information and references on the history and development of metric-affine gravity and known solutions of this theory can be found.
\subsection{Notation}
Our notation follows Refs.~\cite{vassiliev2005quadratic,pasic2014pp,vassiliev2002pseudoinstantons,pasic2010new,pasic2014new,pasic2005pp,king2001torsion,pasic2015torsion}. We denote local coordinates by $x^{\mu}$, where
$\mu = 0, 1, 2, 3$, and write $\partial_{\mu}:=\partial/\partial x^{\mu}.$
We define torsion as
${T}^\lambda{}_{\mu\nu}={\Gamma}^\lambda{}_{\mu\nu}-{\Gamma}^\lambda{}_{\nu\mu}.$
The irreducible pieces of torsion, as given in Ref.~\cite{vassiliev2002pseudoinstantons}, are
$
T^{(1)}=T-T^{(2)}-T^{(3)}$, ${T^{(2)}}_{\lambda\mu\nu}= g_{\lambda\mu}v_\nu-g_{\lambda\nu}v_\mu$,
$T^{(3)}=\ast w$,
where $v_\nu=\frac{1}{3}{T}^{\lambda}{}_{\lambda\nu},\ w_\nu=\frac{1}{6}\sqrt{|\det g|}{T}^{\kappa\lambda\mu}{\varepsilon}_{\kappa\lambda\mu\nu}$.
The irreducible pieces $T^{(1)},T^{(2)}$ i $T^{(3)}$ are called \emph{tensor torsion}, \emph{trace torsion}, and \emph{axial torsion}
respectively.
The interval is $ds^2:=g_{\mu\nu}dx^\mu dx^\nu$.
We say that our connection $\Gamma$ is metric compatible if $\nabla g = 0$. We use the term `parallel' to describe the situation when the covariant derivative of some spinor or tensor field is identically zero.
Given a scalar function $f:M\rightarrow R$ we write
$\int f:=\int f \sqrt{\vert \mathrm{det} g \vert}\mathrm{d}x^0\mathrm{d}x^1\mathrm{d}x^2\mathrm{d}x^3,\ \ \mathrm{det} g:=\mathrm{det}(g_{\mu\nu}).$ We use the spinor formalism introduced in our previous work.\textsuperscript{\cite{pasic2014pp,pasic2005pp}}

\section{Generalising PP-waves to Spacetimes with Axial Torsion}\label{sec2}
In constructing our solutions of the system \eqref{eulerlagrangemetric}, \eqref{eulerlagrangeconnection} we use classical pp-waves, see Refs.~\cite{pasic2014pp,pasic2005pp} and extensive references therein for much more information on pp-waves and known pp-wave type solutions of metric-affine gravity.
We define a \emph{pp-wave} as a Riemannian spacetime which admits a nonvanishing parallel spinor field. Vassiliev showed that pp-waves of parallel Ricci curvature are solutions of the system of equations \eqref{eulerlagrangemetric}, \eqref{eulerlagrangeconnection}.\textsuperscript{\cite{vassiliev2005quadratic}}
We denote the  nonvanishing parallel spinor field by $\displaystyle \chi=\chi^a$ and we assume this spinor field to be \emph{fixed}.
Put
$ l^\alpha:=\sigma^\alpha{}_{a\dot
b}\,\chi^a\bar\chi^{\dot b}
$
where $\sigma^\alpha$ are Pauli matrices.
Then $l$ is a nonvanishing parallel real null vector field. We define the real scalar function
\begin{equation}\label{phase}
\varphi:M\to\mathbb{R},\quad \varphi(x):=\int l\cdot
d x\, ,
\end{equation} which we call the \emph{phase}.
Put
$ F_{\alpha\beta}:=\sigma_{\alpha\beta
ab}\,\chi^a\chi^b
$
where the $\sigma_{\alpha\beta}$ are the `second order Pauli matrices'
\begin{equation}\label{secondPauli}
\sigma_{\alpha\beta ac}:=\frac12
\bigl( \sigma_{\alpha a\dot b}\epsilon^{\dot b\dot d}\sigma_{\beta
c\dot d} - \sigma_{\beta a\dot b}\epsilon^{\dot b\dot
d}\sigma_{\alpha c\dot d} \bigr).
\end{equation}
$F$ can be expressed as
$F=l\, \wedge\, m
$
where $m$ is a complex vector field satisfying $m_\alpha
m^\alpha=l_\alpha m^\alpha=0$, $m_\alpha\overline{m}^\alpha=-2$.
It is known that pp waves can also be defined as
Riemannian spacetimes whose metric can be written locally in the form
\begin{equation}
\label{metric of a pp-wave} d s^2= \,2\,d x^0\,d x^3-(d x^1)^2-(d
x^2)^2 +f(x^1,x^2,x^3)\,(d x^3)^2
\end{equation}
in some local coordinates $(x^0,x^1,x^2,x^3)$. The curvature
tensor $R$ is linear in $f$, i.e.
\begin{equation*}
R_{\alpha\beta\gamma\delta}=
-\frac12(l\wedge\partial)_{\alpha\beta}\,(l\wedge\partial)_{\gamma\delta}f,
\end{equation*}
where
$(l\wedge\partial)_{\alpha\beta}:=l_\alpha\partial_\beta-\partial_\alpha
l_\beta$.
We will restrict our
choice to those coordinates in which
\begin{equation}
\label{explicit l and a} \chi^a=(1,0), \qquad l^\mu=(1,0,0,0),
\qquad m^\mu=(0,1,\mp\rmi,0).
\end{equation}
With such a choice of coordinates the phase (\ref{phase}) becomes
$\varphi(x)=x^3+\mathrm{const}$.
The curvature of a pp-wave can be rewritten
in invariant form, i.e.
\begin{equation}\label{RiemCurv}
R=-\frac12(l\wedge\nabla)\otimes(l\wedge\nabla)f,
\end{equation}
where $l\wedge\nabla:=l\otimes\nabla-\nabla\otimes l$.
The curvature of a pp-wave has the two irreducible pieces, namely
(symmetric) trace-free Ricci and Weyl, see Ref.~\cite{pasic2014new} for more on the properties of classical pp-waves.

Now we wish to generalise the concept of the classical pp-wave to spacetimes with purely axial torsion.
\begin{definition}
\label{definition of a generalised pp-space} A generalised
pp-wave with purely axial torsion is a metric compatible spacetime with pp-metric and torsion
\begin{equation}
\label{define torsion} T :=  *A
\end{equation}
where $A$ is a real vector field defined by $A = k(\varphi) l$ and $k:\mathbb{R}\mapsto \mathbb{R}$ is an arbitrary real function of the phase (\ref{phase}).
\end{definition}
Note that the real vector field $A$ is a plane wave solution of the polarized Maxwell equation $ \displaystyle *d A=\pm\rmi d A$.
It has been suggested, see Refs.~\cite{king2001torsion,vassiliev2000tensor}, that one can interpret the axial component of torsion as the Hodge dual of the electromagnetic vector potential.

We list below the main properties of these generalised pp-waves. Note that here and
further on we denote by $\{\!\nabla\!\}$ the covariant derivative
with respect to the Levi-Civita connection which should not be
confused with the full covariant derivative $\nabla$ incorporating
torsion. We can express torsion as
\begin{equation}\label{torsion2}
T = \mp \frac{i}{2} k(x^3) \ l\wedge m \wedge \overline{m},
\end{equation}
using our special local coordinates \eqref{metric of a pp-wave}, \eqref{explicit l and a}.
Torsion (\ref{define torsion}), (\ref{torsion2}) is clearly purely axial, as $T_{\kappa\mu\nu} = k(\varphi) l^\lambda \varepsilon_{\lambda\kappa\mu\nu}$,  i.e. it is the totally antisymmetric part of torsion.

\begin{remark}
Torsion (\ref{define torsion}), (\ref{torsion2}) corresponds to the axial torsion first obtained by Singh.\textsuperscript{\cite{singh1990axial,singh1990null}} Put $m = -\frac12 k(x^3)$ in formula (16) of Ref.~\cite{singh1990axial} or put $n=0, m = -\frac12 k(x^3)$ in formula (20) of Ref.~\cite{singh1990null}.
\end{remark}
Note that the connection of a generalised pp-wave with purely axial torsion is \emph{metric compatible}, i.e. nonmetricity is absent.
The curvature of a generalised pp-wave has a very nice explicit formula
\begin{eqnarray}
R&=&-\frac12(l\wedge\{\!\nabla\!\})\otimes(l\wedge\{\!\nabla\!\})f \nonumber \\
\label{curvature of a generalised pp-space}
&+&\frac14 k(\varphi)^2 \ \mathrm{Re} \left((l\wedge m)\otimes(l\wedge \overline{m})\right)
\mp\frac12 k'(\varphi) \ \mathrm{Im} \left((l\wedge m)\otimes(l\wedge \overline{m})\right). \label{cuvatureFormula}
\end{eqnarray}
It is a highly nontrivial fact that the torsion generated curvature
i.e.
\begin{eqnarray}
{R_T}^\kappa{}_{\lambda\mu\nu} &=& \partial_\mu{K^\kappa}_{\nu\lambda}
-\partial_\nu{K^\kappa}_{\mu\lambda}
+{K^\kappa}_{\mu\eta}{K^\eta}_{\nu\lambda}
-{K^\kappa}_{\nu\eta}{K^\eta}_{\mu\lambda} \nonumber \\
&=&\frac14 k(\varphi)^2 \mathrm{Re} \left((l\wedge m)\otimes(l\wedge \overline{m})\right)
\mp\frac12 k'(\varphi) \mathrm{Im} \left((l\wedge m)\otimes(l\wedge \overline{m})\right),  \label{torsionCurv}
\end{eqnarray}
and the Riemannian curvature (\ref{RiemCurv}) simply add up to produce formula \eqref{cuvatureFormula}.
Ricci curvature now reads
\begin{equation}\label{Ricci}
Ric = \frac12 (f_{11} + f_{22}-k^2)\  l\otimes l,
\end{equation} where $f_{\alpha\beta} = \partial_\alpha \partial_\beta f$. Note that the scalar curvature is clearly zero and
Ricci curvature is zero if Poisson's equation
$
f_{11} + f_{22} = k
$ is satisfied.
Ricci curvature (\ref{Ricci}) is parallel (i.e. $\nabla Ric = 0$), if
$
f_{11}+f_{22}=k^2 + C,
$ in which case
$Ric = \Lambda \ l\otimes l,$ for some constant $\Lambda$.
\section{New Solutions of Quadratic Metric-affine Gravity}\label{sec3}
Our main result is the following
\begin{theorem}\label{maintheorem}
Generalised pp-waves with purely axial torsion and parallel $\{\!Ric\!\}$ curvature are solutions of the system \eqref{eulerlagrangemetric}, \eqref{eulerlagrangeconnection} in the Yang-Mills case (\ref{YMcase}).
\end{theorem}
Note that by $\{\!Ric\!\}$ we denote the Ricci curvature generated by the Levi-Civita connection only. The condition $\{\!\nabla\!\} \{\!Ric\!\} = 0 $ implies that $f_{11}+f_{22}=C$. Note that the result also holds if $Ric$ is assumed to be parallel.
The torsion generated curvature (\ref{torsionCurv}) has two irreducible pieces of curvature, namely $R^{(1)}$ and $R^{(5)}$, using notation of Ref.~\cite{vassiliev2002pseudoinstantons}, or $^{(2)}W_{\alpha\beta}$ and $^{(4)}W_{\alpha\beta}$, using notation of Ref.~\cite{hehl1995metric}, or $^{(2)}R^{\alpha\beta}$ (Paircom (9)) and $^{(4)}R^{\alpha\beta}$ (Ricsymf (9)), using notation of  Ref.~\cite{blagojevic2013gauge}. The proof of the Theorem \ref{maintheorem} can be found in Ref.~\cite{pasic2015torsion}.

\begin{remark}
It can be shown that for an arbitrary purely axial torsion, the Weyl curvature of the resulting torsion generated curvature is zero.
\end{remark}
Using notation from Refs.~\cite{vassiliev2005quadratic,pasic2014pp,pasic2010new,pasic2005pp}, the irreducible subspaces from which our torsion generated curvature hails are ${\bf R}^{(9,l)}$ and ${\bf R_*}^{(9,l)}$ which can be easily shown by using the explicit formula for the right Hodge dual of torsion generated curvature
\[
{(R_T)^*} = \pm \frac14 k(x^3)^2 \ \textrm{Im }((l\wedge m)\otimes(l\wedge \overline{m}))
+ \frac12k'(x^3) \textrm{Re }((l\wedge m)\otimes(l\wedge \overline{m})).
\]
We want to see whether our spacetimes are solutions of the problem \eqref{eulerlagrangemetric}, \eqref{eulerlagrangeconnection} for the most general quadratic form, which we suspect to be true as well. Hence we propose the following
\begin{proposition}\label{prop1}
Generalised pp-waves with purely axial torsion and parallel $\{\!Ric\!\}$ curvature are solutions of the system of equations \eqref{eulerlagrangemetric}, \eqref{eulerlagrangeconnection} in the most general case.
\end{proposition}
This result is substantially more difficult to show, however, in view of the result from Ref.~\cite{pasic2014pp} in which we showed a similar result using purely tensor torsion and the result from Ref.~\cite{pasic2015torsion} where we proved Theorem \ref{maintheorem}, we hope to be able to prove Proposition \ref{prop1} in the immediate future.
\section{Discussion and Physical Interpretation}
The solution presented in Section \ref{sec2} and in Ref.~\cite{pasic2015torsion} is a completely new solution of quadratic metric-affine gravity and it differs substantially from our previous results obtained in Refs.~\cite{pasic2014pp,pasic2005pp}, where torsion was purely tensor, as well as those from Ref.~\cite{singh1990axial}, where the metric is different and the author only considers the Yang-Mills case. Blagojevi\' c and Hehl provide a review of other exact solutions in this theory.\textsuperscript{\cite{blagojevic2013gauge}}

Similarly to the approach of Refs.~\cite{pasic2014pp,pasic2005pp}, we wish to compare these solutions to the classical model describing the interaction of gravitational and massless neutrino fields, namely Einstein-Weyl theory. Our torsion (\ref{define torsion}) and the corresponding torsion generated curvature (\ref{torsionCurv}) can be considered as waves traveling at speed of light, because $k$ is a function of the phase (\ref{phase}) which has the role of a null
coordinate, $g^{\mu\nu}\nabla_\mu\varphi\,\nabla_\nu\varphi=0$. We choose to deal with the complexified curvature
\begin{equation}\label{complexCurv}
\mathfrak{R}:=r\,(l\wedge m)\otimes(l\wedge \overline{m}),
\end{equation}
where $\displaystyle r:=\frac14 k^2- \frac{i}{2} k'$. This complexification is in line with traditions of quantum mechanics. The complexified curvature (\ref{complexCurv}) is polarized, i.e. ${}^*\mathfrak{R}=\mathfrak{R}^*=\pm\rmi\mathfrak{R}\,,$ and equivalent to a (symmetric) rank $4$ spinor $\omega$, and we have that
$$
\mathfrak{R}_{\alpha\beta\gamma\delta} = \sigma_{\alpha\beta
ab}\,\omega^{abcd}\,\overline{\sigma}_{\gamma\delta cd},
$$ where $\sigma_{\alpha\beta}$ are the second order Pauli matrices (\ref{secondPauli}). Using the properties of pp-waves, we get that $\omega=\xi\otimes\xi\otimes\xi\otimes\xi$, where $\xi:=r^{1/4}\chi$ and $\chi$ is the spinor from Section \ref{sec2}.

Note that our model differs from the classical in that we have conformal invariance, while the classical models contain the gravitational constant. We claim that the spinor field $\xi$ satisfies the massless Dirac equation $\sigma^\mu{}_{a\dot b}\nabla_\mu\,\xi^a=0$, see Ref.~\cite{pasic2014pp} for more details. Furthermore, as was similarly done in Ref.~\cite{pasic2014pp}, we can construct pp-wave type solutions of Einstein-Weyl theory very similar to purely axial torsion pp-wave solutions of QMAG from Section \ref{sec3} to propose that generalised pp-waves with purely axial torsion represent a metric-affine model for the massless neutrino.

\section*{Acknowledgments}
We would like to thank the organisers of the 14th Marcel Grossmann meeting for awarding us grants. We would also like to thank Dmitri Vassiliev, Milutin Blagojevi\' c and Friedrich Hehl for helpful advice. Special thanks goes to Prof. Izudin Kapetanovi\' c for all his kind assistance, as well as Solana d.d. Tuzla.

\bibliographystyle{MyBST}
\bibliography{bibliography}

\label{lastpage}

\end{document}